%% file: paper.tex
\documentclass[twocolumn,showpacs,aps,prl,superscriptaddress]{revtex4}

\usepackage{graphicx}
\usepackage{dcolumn}
\usepackage{amsmath}
\usepackage{epsfig}

\input babarsym

\newcommand{\BABARPubYear}    {08}
\newcommand{\BABARPubNumber}  {007}

\newcommand{\SLACPubNumber} {13179}

\newcommand{\lumi}    {413\invfb}
\newcommand{\nBB}     {455\times 10^6}

\def\figurebox#1#2#3{%
    \def\arg{#3}%
    \ifx\arg\empty
    {\hfill\vbox{\hsize#2\hrule\hbox to #2{\vrule\hfill\vbox to #1{\hsize#2\vfill}\vrule}\hrule}\hfill}%
    \else
    {\hfill\epsfbox{#3}\hfill}%
    \fi}

\begin{document}

\def\Xz {\ensuremath{X(3872)}\xspace}
\def\mES {\ensuremath{m_{\mbox{\scriptsize{ES}}}}\xspace}
\def\BB {\ensuremath{B\bar{B}}\xspace}

\preprint{\babar-PUB-\BABARPubYear/\BABARPubNumber} 
\preprint{SLAC-PUB-\SLACPubNumber} 

\begin{flushleft}
\babar-PUB-\BABARPubYear/\BABARPubNumber\\
SLAC-PUB-\SLACPubNumber\\
%hep-ex/\LANLNumber\\[10mm]
\end{flushleft}

\title{
{\large \bf Study of \boldmath \B \to \Xz $K$, with \Xz \to \jpsi $\pip$ $\pim$ 
} 
}

% Dummy author list; contact PubBoard Chair for current author list
%\input authors

\input authors_feb2008.tex

\date{\today}% It is always \today, today, but you may specify any date with \date.

\begin{abstract}
We present measurements of the decays $\Bp \to \Xz \Kp$ and $\Bz \to \Xz \Kz$
with $\Xz\to\jpsi\pip\pim$. The data sample used, collected with 
the \babar\ detector at the PEP-II $e^+e^-$ asymmetric-energy storage ring, corresponds 
to $\nBB B\bar{B}$ pairs. Branching fraction measurements of
${\cal B}(\Bp \to \Xz \Kp) \times {\cal B}(\Xz\to\jpsi\pip\pim) = (8.4\pm 1.5 \pm 0.7) \times 10^{-6}$ 
and  
${\cal B}(\Bz \to \Xz \Kz) \times {\cal B}(\Xz\to\jpsi\pip\pim) = (3.5\pm 1.9 \pm 0.4) \times 10^{-6}$ 
are obtained. We set an upper limit on the natural width of the $\Xz$ of 
$\Gamma < 3.3 \mevcc$ at the 90\% confidence level.

\end{abstract}

% PACS, the Physics and Astronomy Classification Scheme.

\maketitle
%%
%% --------- Introduction ----------------
%%
The X(3872) was discovered in 2003 by the Belle Collaboration 
\cite{BelleDiscovery} which reported the observation of a narrow resonance in exclusive \Bpm decays to 
\Kpm (\jpsi \pip \pim).  The new state was then confirmed by CDF \cite{cdfXconfirmation}, 
D0 \cite{d0Xconfirmation}, and \babar\ \cite{babarXfirst, babarXsecond}. 

There has been great interest in this narrow state, with numerous theoretical interpretations
having been proposed, including a $\bar{D^0} D^{*0}$ molecule, a diquark-antidiquark, a tetraquark 
state, a hybrid charmonium or a classical charmonium state. 
The diquark-antidiquark model~\cite{Maiani:2004vq} predicts the \Xz states 
to be produced at equal rates in \Bz and \Bp decays with a mass difference of $(8\pm 3)\mevcc$. 
The $S$-wave molecule model \cite{swavemolecule} predicts the neutral $B$ branching fraction 
to be much smaller than the charged $B$ one. 

Studies of angular distributions by CDF \cite{CDFXAngular} 
favor the quantum number assignment $J^{PC} = 1^{++} \ {\rm or}\ 2^{-+}$.

The X(3872) has also been observed in the $\Xz \to \jpsi \gamma$ decay mode by  
\babar\ \cite{BabarXjpsigamma}, indicating that it must have positive $C$-parity. 
Therefore, the \pip\pim pair in the $\Xz \to \jpsi \pip \pim$ must have a negative $C$-parity and an 
odd orbital angular momentum $L$, to satisfy $C$(\pip\pim)=$-$1=$(-1)^{L+S}$, with $S=0$. 
The \pip\pim invariant 
mass distribution has been studied by CDF \cite{CDFpipi}, and 
found to be consistent with a $\rho^0$ meson, where the \jpsi and the $\rho^0$ are in a relative 
$S$-wave. 

Both \babar\ \cite{babard0d0bar} and Belle \cite{belled0d0bar} have observed the $\Xz \to \bar{D^0}D^{*0}$ 
decay. These searches were motivated by the fact that the X(3872) was barely above the $\bar{D^0}D^{*0}$ 
mass threshold. The mass measurement results are very consistent between the two experiments but are about 
3 \mevcc (representing $\simeq 4$ standard deviations) above the mass measured in the $\jpsi \pip \pim$ 
decay mode.

We report herein an analysis of $\Bp \to \Xz \Kp$ and $\Bz \to \Xz \Kz$, 
with $\Xz \to \jpsi \pip \pim$, where charge conjugation is implied throughout. 
We present updated branching fractions for 
the two channels and extract the mass and width of the X(3872) state.
%%
%% --------- Lumi ----------------
%%
The data sample used for this analysis, collected 
by the \babar\ detector at the \pep2\ asymmetric-energy \epem\ 
storage ring operated at the Stanford Linear Accelerator Center, corresponds to a
total integrated luminosity of 413 fb$^{-1}$, recorded at the \FourS resonance.

%%
%% --------- Detector ----------------
%%
The \babar\ detector is described in detail in Ref.~\cite{detector}.
Charged particle  momenta are measured with a 5-layer
double-sided silicon vertex tracker (SVT) and a 40-layer drift chamber (DCH) 
inside a 1.5-T superconducting solenoidal magnetic field.
A calorimeter consisting of 6580 CsI(Tl) 
crystals (EMC) is used to measure electromagnetic energy. 
A ring-imaging Cherenkov detector (DIRC) is used to identify
charged hadrons, aided by the $dE/dx$ measurement in SVT and DCH.
Muons are identified by the instrumented magnetic flux return (IFR).
Particle attributes are reconstructed in the laboratory
frame and then boosted to the \epem\ center-of-mass (CM) 
frame using the asymmetric beam energy information.
%%
%% --------- MC-Sample ----------------
%%
We use a GEANT~4~\cite{geant4} Monte Carlo simulation (MC) to estimate the signal efficiencies, 
employing a sample in which one of the generated $B$ mesons decays to the 
signal mode and the other to a representative sample of $B$ decays.

%%
%% --------- Analysis Overview ----------------
%%

Charged particles are required to have transverse momenta greater than 100 MeV/$c$ in the laboratory 
frame. The distance of closest approach of charged tracks must be within $\pm 10$ cm
of the $z$ coordinate (along the beam axis) of the primary vertex and within a
circle of radius 1.5 cm in the $x-y$ plane. Kaons, electrons and muons are separated from pions
based on information from the IFR and DIRC, energy loss in the SVT and DCH ($dE/dx$), and the ratio 
of the associated EMC energy deposition ($E_{\rm cal}$) to its momentum ($E_{\rm cal}/p$). 

The $\Bp\to\jpsi\pip\pim\Kp$ and $\Bz\to\jpsi\pip\pim\KS$ decays are
reconstructed as follows. Electrons and
bremsstrahlung photons satisfying $2.95<m(\epem(\gamma))<3.14~\gevcc$
are used to form $\jpsi\to\epem$ candidates.  A pair of muons 
within the mass interval $3.06<m(\mumu)<3.14~\gevcc$ is
required for a $\jpsi\to\mumu$ candidate. A mass constraint to the
nominal \jpsi mass~\cite{pdg} is imposed in the fit of the lepton
pairs.  We reconstruct $\KS\to\pip\pim$ candidates from pairs of
oppositely charged tracks forming a vertex with a $\chi^2$ probability
greater than $0.1\%$, a flight-length ($l$) significance $l/\sigma(l)>16$ 
(where $\sigma(l)$ is the measurement error) and 
an invariant mass within $15~\mevcc$ of the nominal $\KS$
mass~\cite{pdg}. We form $\Xz$ candidates by combining \jpsi
candidates with two oppositely charged pion candidates, all fitted to a
common vertex. Finally, we form $\Bp(\Bz)$ candidates by combining
$\Xz$ candidates with $\Kp(\KS)$ candidates. To suppress continuum
background, we select only events with a ratio of the second to the zeroth
Fox-Wolfram moment~\cite{Fox1979} less than $0.5$.

We use two kinematic variables to identify signal events coming from $B$ decays: the
difference between the energy of the $B$ candidate and the beam energy,
$\DeltaE = E_B^*-\sqrt{s}/2$, and the energy-substituted mass
$\mes = \sqrt{(s/2+{\bf p}_i\cdot{\bf p}_B)^2/E_i^2 - {\bf p}_B^2}$. Here
$(E_i,{\bf p}_i)$ is the four-vector (in the laboratory frame) and $\sqrt{s}$ is the
center-of-mass (CM) energy of the $\epem$ system, $E_B^*$ is the energy of the
$B$ candidate in the CM system and ${\bf p}_B$ the momentum in the
laboratory frame. The signature of signal events is $\Delta E\approx 0$, and $\mes
\approx m_{B}$, where $m_{B}$ is the nominal mass of the \hbox{$B$ meson}~\cite{pdg}.

%%
%% --------- Selection ----------------
%%

If there are multiple candidates in a single event (about $9\%$ of the events), we select the candidate
with the smallest value of $|\DeltaE|$. 
We optimize the signal selection criteria by maximizing the ratio
$n_{\rm{s}}^{\rm{mc}}/(a/2+\sqrt{n_{\rm{b}}^{\rm{mc}}})$~\cite{Punzi:2003bu}, where $a=3$ represents 
the desired significance of signal-to-background ratio in number of sigmas  
and $n_{\rm{s}}^{\rm{mc}}$ ($n_{\rm{b}}^{\rm{mc}}$) are the number of reconstructed Monte Carlo signal
(background) events. The optimization is performed by varying the selected ranges of \DeltaE, 
$|\mes-m_B|$, \Xz and \KS candidate masses, the \KS\ flight significance and the particle 
identification (PID) selection requirements for the leptons, pions and charged kaons. 
The criteria $|\DeltaE|<20~\mev$ and $|\mes-m_B|<6~\mevcc$, which 
represent  about three standard deviations of the resolution of the 
quantities, were found to be optimal for selecting signal events.

We extract the number of signal events with an extended, unbinned
maximum-likelihood fit to the $m_X$ distribution, where $m_X$ is the 
$\jpsi\pip\pim$ invariant mass. 

The probability density function (PDF), normalized to the total number of events, 
is $\mathcal{P}(m_X) = \sum_{t}n_t\mathcal{P}_t(m_X)$ where $n_t$
is the number of events of category $t$ and $\mathcal{P}_t$ is the associated PDF. 
We consider only two different
event categories: signal and combinatorial background (which arises mainly from $B$ decays). 
The signal PDF is modelled by a Lorentzian function that describes both the natural width 
and the experimental resolution. We model combinatorial background events by a linear function in $m_X$. 
For the neutral mode fit, 
the width of the Lorentzian has been fixed to the value obtained from the charged 
mode fit.

\begin{figure}[ht]
\begin{center}
  \includegraphics[width=0.48\textwidth]{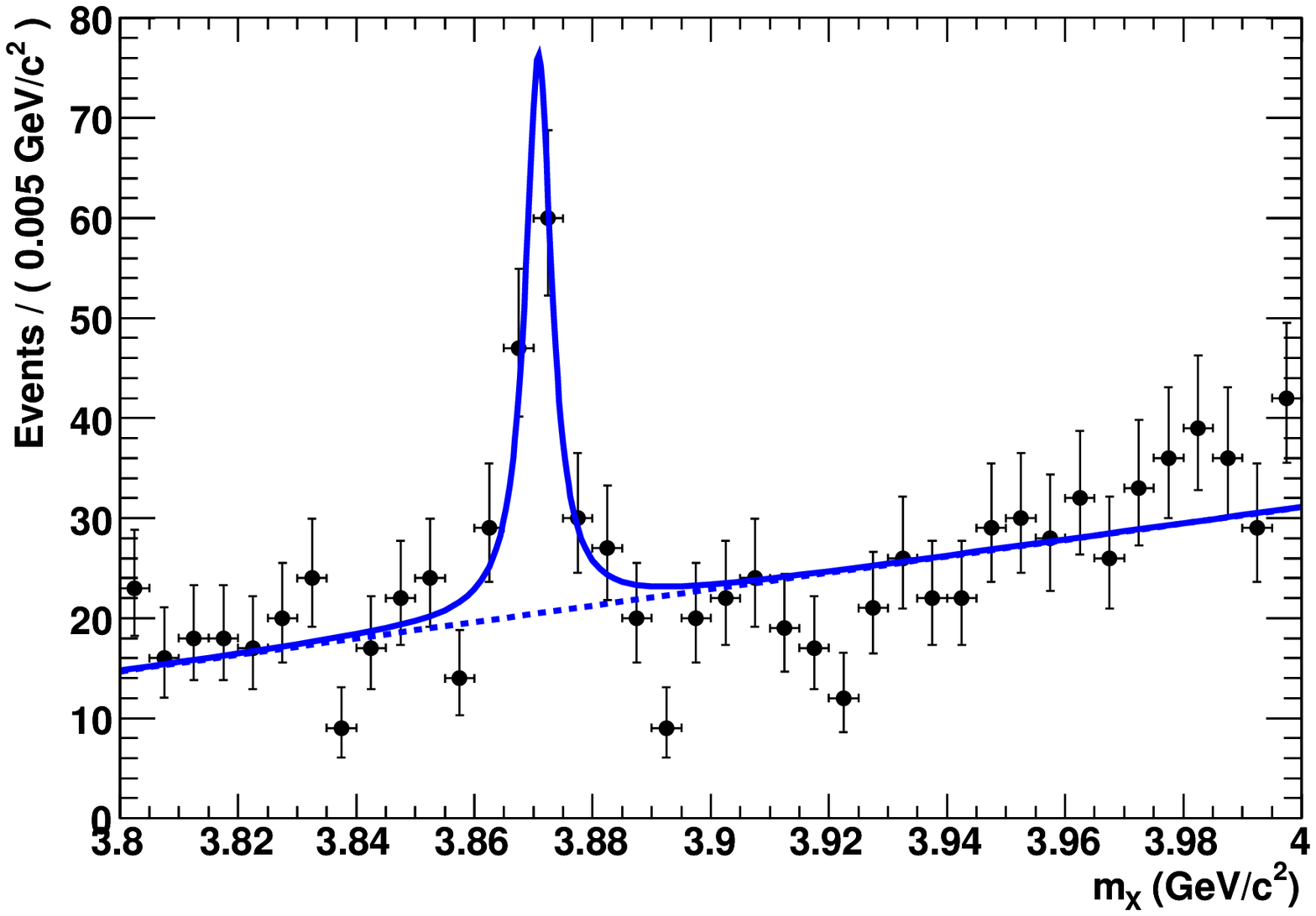}
  \includegraphics[width=0.48\textwidth]{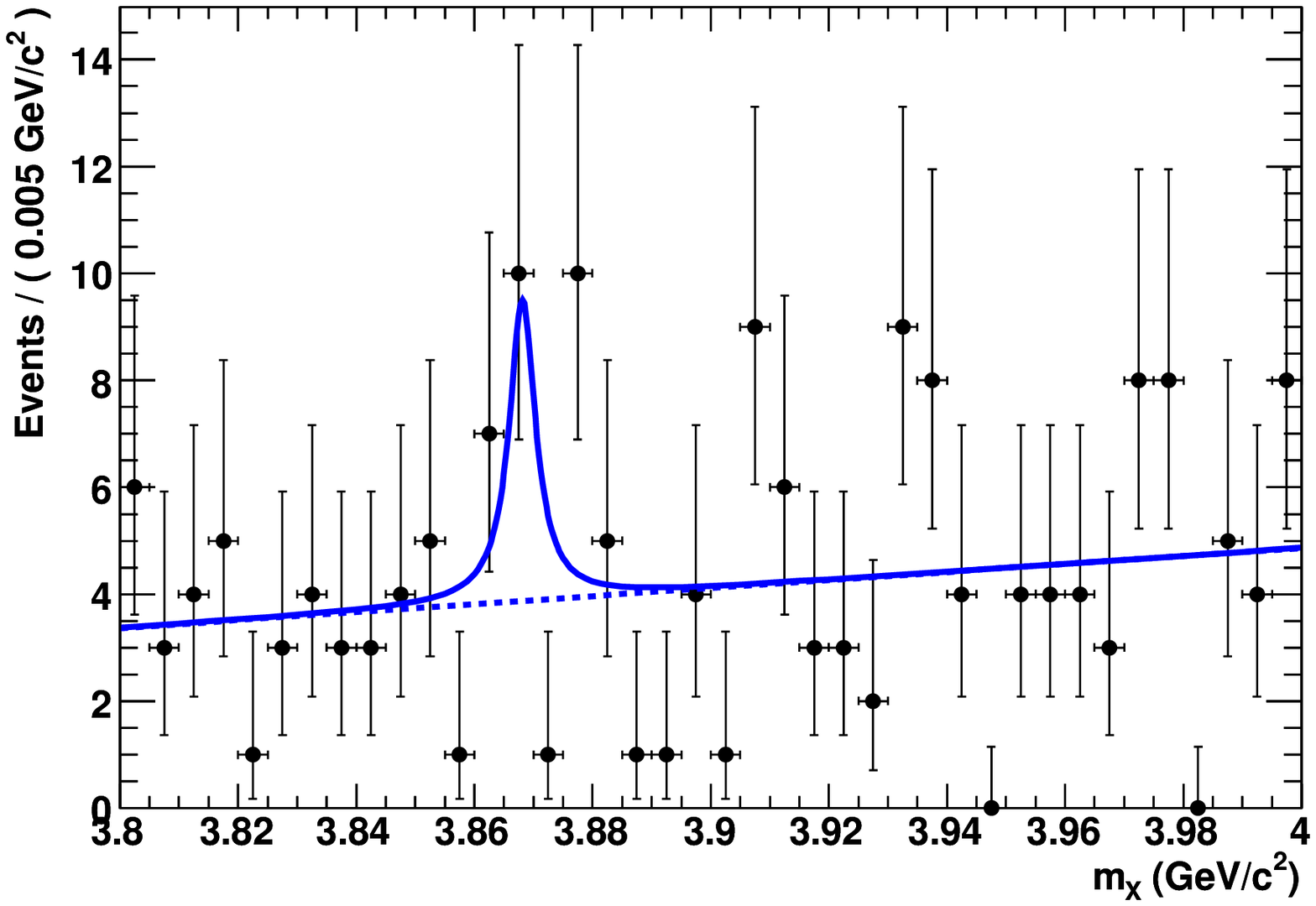}
  \caption{\label{fig:fit} 
  Fits to the $m(\jpsi\pip\pim)$ distributions of (top) $\Bp\to\Xz\Kp$ 
  and (bottom) $\Bz\to\Xz\KS$ candidates drawn from the \lumi sample. 
  The dashed line represents the combinatorial background PDF
  and the solid line the sum of background plus the signal
  PDF. }
\end{center}
\end{figure}

The fit is performed in the region $3.8<m_X<4.0~\gevcc$, after applying 
the optimized selection criteria on all other variables. 
The signal region projections of the
one-dimensional fit to the data are shown in Fig.~\ref{fig:fit} for the $B^+$
(top) and $B^0$ (bottom) modes.
We obtain $93.4\pm17.2$ signal events for the $\Bp$ mode ($n_{\rm{s}}^+$) and
$9.4\pm 5.2$ signal events for the $\Bz$ mode ($n_{\rm{s}}^0$).
We interpret the observed events in either mode as the $X(3872)$.
Results are summarized in Table~\ref{tab:FitResultsDataX}. We fit the 
$\jpsi\pip\pim$ system invariant mass in the \mes\ side band region ($\mes < 5.27~\gevcc$) and 
observe no signal.

%%
%% --------- Efficiency Method ----------------
%%

The efficiency is determined from MC samples with a \Xz
signal of zero natural width at $3.872~\gevcc$. The decay model consists of the
sequential isotropic decays $B\to\Xz K$, $\Xz\to\jpsi\rho^0$, and
$\rho^0\to\pip\pim$. This yields a more accurate description of the observed $\pip\pim$ invariant mass
distribution~\cite{babarXfirst}, compared to a three-body decay.  Efficiencies are corrected for the
small differences in \KS\ reconstruction efficiencies that are found by
comparing data and MC control samples. The final reconstruction efficiencies are
$(20.60\pm0.10)\%$ for the $\Bp\to\Xz\Kp$ mode and $(14.50\pm0.09)\%$ for the $\Bz\to\Xz\KS$   
mode, where the errors are dominated by the size of the signal MC samples.

\begin{table}[htbp]
\caption{\label{tab:FitResultsDataX}
Fit results for both $\Bp\to\Xz\Kp$ and $\Bz\to\Xz\KS$ modes. 
In the fit to the \Bz mode, the width is fixed to the value obtained from 
the \Bp mode. Errors are statistical only. The mass measurements are subsequently 
corrected.}
\begin{center}
\begin{small}
\begin{tabular}{l|r|r}
\hline
\hline
 &  $\Bp\to\Xz\Kp$ & $\Bz\to\Xz\KS$ \\ \hline
{\bf Parameters}  & & \\
$m_{X,\text{fit}}$                    & $3870.86\pm0.60$  & $3868.13\pm1.53$  \\
$m_X$ Lorentz $\Gamma_{\rm fit}$         & $5.43\pm1.52$     & Fixed to 5.43     \\ 
Linear Background              &                   &                   \\
slope                          & $ -0.30\pm0.04$   & $-0.28\pm0.03$    \\ \hline
{\bf Yields}  & & \\
signal $n_S$  & $ 93.4\pm17.2$ & $ 9.4\pm5.2$ \\ \hline\hline
\end{tabular}
\end{small}
\end{center}
\end{table}

%%
%% --------- Systematics Results ----------------
%%
The fit is validated with MC experiments, where 
we embed samples of the number of expected signal events into 
MC background samples. On average, the number of signal events found is 
in good agreement with the signal sample size. The fit is further validated with a set 
of parameterized MC experiments, based on the signal PDF parameters, which return 
the number of input signal events with no significant bias.

The systematic errors on the branching fraction are summarized in  Table~\ref{tab:summarysyste}. 
They include uncertainties in the number of \BB events, secondary
branching fractions ~\cite{pdg}, efficiency calculations due 
to limited MC statistics, the MC decay model of the \Xz, PID, 
charged particle tracking, \KS reconstruction, background modelling (BM) and those arising 
from fixing the width in the $\Bz\to\Xz\KS$ mode. The production ratio of \Bz and \Bp 
mesons in \FourS decays is taken to be $1.031\pm0.033$~\cite{pdg}. 

The total fractional errors, $8.8\%$ and $11.7\%$ for the $\Bp$ and $\Bz$ modes, respectively, are 
obtained by adding the uncertainties in Table \ref{tab:summarysyste} in quadrature.

\begin{table}[htbp]
\caption{\label{tab:summarysyste}
Summary of (fractional) systematic uncertainties on the branching fraction measurements for 
both modes.}
\begin{center}
\begin{small}
\begin{tabular}{l|r|r} \hline \hline
                     & $\Bp\to\Xz\Kp$  &  $\Bz\to\Xz\KS$\\ \hline
Tracking       & 1.8 &  1.4 \\
\KS correction & n/a &  0.7 \\
PID            & 1.9 &  1.4 \\
MC Model       & 1.3 &  0.9 \\
$B$ counting     & 1.1 &  1.1 \\
MC statistics  & 0.5 &  0.6 \\
Secondary BF   & 3.3 &  3.3 \\ 
BM             & 7.5 &  4.2 \\ 
Fixed width    & n/a & 10.1 \\ \hline
Total Fractional Error         & 8.8 & 11.7 \\ \hline \hline
\end{tabular}
\end{small}
\end{center}
\end{table}

The significance is estimated as $\sqrt{-2\ln(\mathcal{L}_0/\mathcal{L}_{\rm{max}})}$ where 
$\mathcal{L}_{\rm{max}}$ and $\mathcal{L}_{0}$ are likelihoods returned by the nominal 
fit and by the fit with the signal yield fixed at zero. 
The estimated statistical significance  of each 
signal is $8.6\sigma$ and $2.3\sigma$, for the \Bp and \Bz modes respectively.

Using $n^0_{\rm{s}}$ and $n^-_{\rm{s}}$, the efficiencies, the secondary branching
fractions and the number of \BB events, we obtain the branching
fractions
$\BR(\Bz\to\Xz\Kz)\times\BR(X\to\jpsi\pip\pim)=(3.5\pm1.9 \pm0.4)\times 10^{-6}$
and
$\BR(\Bp\to\Xz\Kp)\times\BR(X\to\jpsi\pip\pim)=(8.4\pm1.5 \pm0.7)\times 10^{-6}$. 
We also calculate  a $90\%$ confidence level (C.L.) upper limit on the neutral branching fraction 
as $\BR(\Bz\to\Xz\Kz)\times\BR(X\to\jpsi\pip\pim)<6.0 \times 10^{-6}$ ($90\%$, C.L.).
For the ratio of branching fractions, in which most of the systematic errors cancel, we obtain
\begin{eqnarray}
  \label{eq:R}
    \nonumber
   R(X) = 
   \frac{\BR(\Bz\to\Xz\Kz)}{\BR(\Bp\to\Xz\Kp)} 
   = 0.41\pm0.24 \pm0.05,
\end{eqnarray}
where the first(second) uncertainty is statistical(systematic). Assuming 
Gaussian errors, we calculate the upper limit $R(X)<0.73$ at $90\%$ C.L. 

We use the \Bp\to\psitwos\Kp and \Bz\to\psitwos\Kz decays, with $\psitwos\to\jpsi\pip\pim$, 
in the \psitwos mass region~\cite{pdg} 
as control modes. We measure the branching fractions and obtain the ratio of 
neutral to charged branching fractions 
$R(\psitwos) = 0.81\pm0.05\pm0.01$, in agreement with the world average of $0.96\pm0.11$.
We also use these control modes to correct the \Xz mass. 
We fit the $\jpsi\pip\pim$ invariant mass in the \psitwos and \Xz region. 
We correct the \Xz mass measurement, $m_{X,\rm{fit}}$, by the difference between the \psitwos world 
average mass~\cite{pdg}, $m_{\psitwos}$, and its measured mass, $m_{\psitwos,\text{fit}}$, which yields 
$m_X = m_{X,\text{fit}}-m_{\psitwos,\text{fit}}+m_{\psitwos}$. 
The result for the \Bp mode is $(3871.4\pm0.6 \pm0.1)\mevcc$ and $(3868.7\pm1.5 \pm0.4)\mevcc$
for the \Bz mode, where the first error is the statistical uncertainty on $m_{X,\text{fit}}$ 
and the second is the uncertainty on $m_{\psitwos,\text{fit}}$ and $m_{\psitwos}$. In the 
neutral mode we have also included an uncertainty that arises from fixing the width to the value obtained 
from the charged mode. 
The mass difference of the \Xz states produced in \Bz and \Bp decays is 
$\Delta m=(2.7\pm 1.6\pm 0.4)\mevcc$.

The natural width $\Gamma_{\rm X}$ of the \Xz is obtained using the \Bp mode by subtracting the full width 
at half maximum of the resolution function measured from Monte Carlo $\Gamma_{\rm Res}$ from 
the data $\Gamma_{\rm fit}$, $\Gamma_{\rm X} = \Gamma_{\rm fit} - \Gamma_{\rm Res}$, with 
both the resolution and the natural width of the \Xz parameterized by 
a Lorenztian function. 
We estimate a systematic error 
on the width by comparing the nominal value to the value determined when using a 
two-Gaussian resolution function. We determine the natural 
width of the \Xz to be $(1.1\pm1.5\pm0.2)~\mevcc$, where the first uncertainty is statistical 
and the second systematic.
From this result we calculate the $90\%$ C.L. upper limit on the natural width  $\Gamma_{\rm X}<3.3~\mevcc$.

%%
%% --------- Summary ----------------
%%

In summary, we have performed an updated study of the decays $\Bp \to \Xz \Kp$ and 
$\Bz \to \Xz \Kz$ with $\Xz\to\jpsi\pip\pim$. The branching fraction measurements in the 
$B^+$ and $B^0$ modes are in good agreement with previous results,
with comparable or better errors. 
The ratio of the branching fractions is $R=0.41\pm0.24 \pm0.05$ and the 
observed mass difference is $\Delta m=(2.7\pm 1.6\pm 0.4)~\mevcc$, consistent 
with either the molecular or diquark-antidiquark model within two standard deviations. 
Finally, we provide an updated upper limit of the natural width of 
the \Xz, $\Gamma_{\rm X}<3.3~\mevcc$ ($90\%$, C.L.).

% Input the pubboard acknowledgements file
\input pubboard/acknowledgements.tex

\end{document}

%% file: authors_feb2008.tex
%% author list as of 05-Feb-2008 (544 authors)
%
\author{B.~Aubert}
\author{M.~Bona}
\author{Y.~Karyotakis}
\author{J.~P.~Lees}
\author{V.~Poireau}
\author{E.~Prencipe}
\author{X.~Prudent}
\author{V.~Tisserand}
\affiliation{Laboratoire de Physique des Particules, IN2P3/CNRS et Universit\'e de Savoie, F-74941 Annecy-Le-Vieux, France }
\author{J.~Garra~Tico}
\author{E.~Grauges}
\affiliation{Universitat de Barcelona, Facultat de Fisica, Departament ECM, E-08028 Barcelona, Spain }
% ...
\collaboration{The \babar\ Collaboration}
\noaffiliation

%% file: pubboard/acknowledgements.tex
We are grateful for the 
extraordinary contributions of our \pep2\ colleagues in
achieving the excellent luminosity and machine conditions
that have made this work possible.
The success of this project also relies critically on the 
expertise and dedication of the computing organizations that 
support \babar.
The collaborating institutions wish to thank 
SLAC for its support and the kind hospitality extended to them. 
This work is supported by the
US Department of Energy
and National Science Foundation, the
Natural Sciences and Engineering Research Council (Canada),
the Commissariat \`a l'Energie Atomique and
Institut National de Physique Nucl\'eaire et de Physique des Particules
(France), the
Bundesministerium f\"ur Bildung und Forschung and
Deutsche Forschungsgemeinschaft
(Germany), the
Istituto Nazionale di Fisica Nucleare (Italy),
the Foundation for Fundamental Research on Matter (The Netherlands),
the Research Council of Norway, the
Ministry of Education and Science of the Russian Federation, 
Ministerio de Educaci\'on y Ciencia (Spain), and the
Science and Technology Facilities Council (United Kingdom).
Individuals have received support from 
the Marie-Curie IEF program (European Union) and
the A. P. Sloan Foundation.